\input phyzzx.tex

\twelvepoint


\def\m{\hat m}
\def\n{\hat n}
\def\p{\hat p}
\def\q{\hat q}
\def\r{\hat r}
\def\s{\hat s}


\REF\HSW{P.S. Howe, E. Sezgin and P.C. West, Phys. Lett. {\bf B399}
(1997) 49, hep-th/9702008}
\REF\others{M. Perry and J.H. Schwarz, Nucl. Phys. {\bf B489} (1997) 47,
hep-th/9611065; M. Aganagic, J. Park, C. Popescu, and J. H. Schwarz,
{\it Worldvolume action of the M-theory fivebrane},
hep-th/9701166;
I. Bandos, K Lechner, A. Nurmagambetov, P. Pasti and D. Sorokin, and M.
Tonin, {\it Covariant action for the super fivebrane of M-theory},
hep-th/9701149}
\REF\PKT{P.K. Townsend, {\it $P$-brane Democrary}, hep-th/9507048}
\REF\AGIT{J.A. de Azcarraga, J.P. Gauntlett, J.M. Izquierdo and  
P.K. Townsend, Phys. Rev. Lett. {\bf 63} (1989) 2443}
\REF\Strominger{A. Strominger, Phys. lett. {\bf B383} (1995) 44, 
hep-th/9512059}
\REF\PT{G. Papadopoulos and P.K. Townsend, Phys, Lett. {\bf B384} 
(1996) 86, hep-th/9605146}
\REF\Polchinski{J. Polchinski, Phys. Rev. Lett. {\bf 75} (1995) 4724, 
hep-th/9510017}
\REF\GH{O. Ganor and A. Hanany, Nucl. Phys. {\bf B474} (1996) 122, 
hep-th/9602120}
\REF\HLW{P.S. Howe, N.D. Lambert and P.C. West, {\it The Self-Dual String 
Soliton}, hep-th/9709014}
\REF\Witten{E. Witten, {\it Solutions of Four Dimensional Field Theories
via M Theory}, hep-th/97003166}
\REF\HLWnew{P.S. Howe, N.D. Lambert and P.C. West, {\it Classical M-Fivebrane
Dynamics and Quantum $N=2$ Yang-Mills}, KCL-TH-97-55}
\REF\Harvey{J.A. Harvey, {\it Magnetic Monopoles, Duality and Supersymmetry}, 
\ \ \ \  hep-th/9603086}

\pubnum={KCL-TH-97-54\cr hep-th/9710033}
\date{September 1997}

\titlepage

\title{\bf The Threebrane Soliton of the M-Fivebrane}

\centerline{P.S. Howe}
\centerline{N.D. Lambert}

\centerline{and}

\centerline{P.C. West\foot{phowe, lambert, pwest@mth.kcl.ac.uk}}
\address{Department of Mathematics\break
         King's College, London\break
         England\break
         WC2R 2LS\break
         }

\abstract
We discuss the supersymmetry algebra of the M theory fivebrane and 
obtain a new threebrane soliton preserving half of the 
six-dimensional supersymmetry. This solution is dimensionally reduced to
various D-$p$-branes.

\endpage


\chapter{Introduction}

The dynamics of the M theory fivebrane are given by an 
interacting  $(2,0)$ tensor multiplet 
containing a self-dual three tensor $h_{mnp}$, ($m,n,p=0,\ldots,5$) 
five scalars 
$X^{b'}$ ($b'=1',\dots,5'$) and sixteen fermions $\Theta_{\alpha}^{\ i}$, 
($\alpha=1,\ldots,4$, $i=1,\ldots 4$). In this paper we shall use the 
formalism and notations of [\HSW] but there are also other
formulations available [\others]. In this short paper we 
wish to examine some properties of the supersymmetric soliton 
states in this theory.

The most general form for the $(2,0)$ supersymmetry algebra in six 
dimensions is
$$
\{Q_{\alpha}^{\ i},Q_{\beta}^{\ j}\} = 
\eta^{ij}(\gamma^m)_{\alpha\beta}P_m 
+ (\gamma^m)_{\alpha\beta}Z^{ij}_m
+ (\gamma^{mnp})_{\alpha\beta}Z^{ij}_{mnp}\ .
\eqn\algebra
$$
Here $\eta^{ij}$ is the invariant tensor of $USp(4)\cong Spin(5)$, 
$P_m$ is the momentum, $Z_m^{ij}$ is in the symmetric ${\bf 5}$ of $Spin(5)$
and $Z_{mnp}^{ij}$ is self-dual and in the anti-symmetric $\bf 10$ of 
$Spin(5)$. 
It is possible to add to \algebra\ a five form central charge in the 
${\bf 5}$ of $Spin(5)$, however, due
to the self-duality constraint it contains no additional degrees of freedom.
In fact the left hand side can be thought of as a sixteen by sixteen 
symmetric matrix and therefore has ${16\times 17 \over 2}=136$ degrees of 
freedom. Similarly the right hand side contains 
$6 + 5\times 6 + 10\times 10 = 136$ 
degrees of freedom and hence there can be no additional independent terms
[\PKT].
Since the $p$-form charges of $p$-dimensional 
extended objects can give rise the additional central charges in \algebra\
[\AGIT], we may expect to find 
onebrane and threebrane solitons of the M-fivebrane equations
preserving half of the spacetime supersymmetry. We will shortly  return to the
interpretation of the five form central charge.

There are in fact two other arguments which also lead to the 
appearance of onebranes and threebranes on the M-fivebrane worldvolume. The 
first is from the M theory interpretation in eleven dimensions. Here one
has two possible configurations which preserve a quarter of the 
eleven-dimensional supersymmetry; a membrane intersecting a fivebrane over a 
onebrane [\Strominger] or two
fivebranes intersecting over a threebrane [\PT]. In terms of the worldvolume of
the fivebrane these configurations 
will appear as onebrane and threebrane solitons
respectively, preserving half of the six-dimensional supersymmetry.

Another reason one may expect only one form and three form central 
charges comes from the interpretation of these configurations as the
effective field outside of a D-$p$-brane [\Polchinski] of the self-dual
string theory. Here we will assume that the notion of a
D-$p$-brane (as a $(p+1)$-dimensional hyperplane in spacetime where open
strings end) can be 
extended to an analogous object of the six-dimensional 
self-dual string theory. In this case the
supersymmetries preserved by the modified boundary conditions on the
self-dual string are those for which
$$
\epsilon_L = \Gamma_0\ldots\Gamma_p\epsilon_R \ ,
\eqn\Dsusys
$$
where the D-brane lies in the $x^0,\ldots,x^p$ plane and $\epsilon_{L}$ and 
$\epsilon_{R}$
are the left and right handed  supersymmetry generators on the self-dual
string worldsheet respectively. Since $\epsilon_L$ and $\epsilon_R$ 
are of the same six-dimensional chirality it is
easy to check that \Dsusys\ is consistent if and only if $p=1,3,5$. In this
case there again appears the possibility of a D-fivebrane on the worldvolume
and hence a five form central charge in \algebra. This
would be analogous the the D-ninebrane of the type IIB string and carry
no physical degrees of freedom. It does however point to the possibility
of performing an orientifold projection to a self-dual $(1,0)$ string
theory in six dimensions, in analogy with the construction of the type
I string as an orientifold of the type IIB string [\Polchinski].
The analogy with type IIB string theory also suggests that the six-dimensional
self-dual string may possess a sort of D-instanton or $-1$-brane. However 
we shall see below that this is not the case.

This $(1,0)$ string theory can be understood within M theory if we  suppose 
that there is some kind of ninebrane. It would then be
possible for a ninebrane to intersect a fivebrane over a fivebrane (i.e.
a fivebrane-ninebrane bound state).\foot{In fact some state of this kind must
exist since there is a type IIA eightbrane-fourbrane bound state preserving
eight supersymmetries, indeed 
we shall construct it below.} This would give rise to a five form
central charge on the fivebrane. The ninebrane would then have four
directions transverse to the fivebrane, indicated by four of the five 
scalars. By choosing which four scalars of the five on the worldvolume to use 
one obtains a multiplet of fivebranes in the $\bf 5$ of 
$Spin(5)$, in agreement with the algebra \algebra. This construction of the
$(1,0)$ self-dual string has appeared in [\GH], in connection with the
fivebrane of the $E_8\times E_8$ heterotic string.

In [\HLW] a family of
supersymmetric 
onebrane solitons on the M-fivebranes equations of motion were found, 
transforming under the ${\bf 5}$ of $Spin(5)$. In this paper we shall seek
a family of supersymmetric threebrane solitons in the ${\bf 10}$ of $Spin(5)$
and thereby complete the $p$-brane spectrum of the  M-fivebrane. 
The threebrane of the M-fivebrane is also
of some interest in its own right due to its relation to the low energy 
Seiberg-Witten effective action of 
$N=2$, $D=4$ Yang-Mills theory [\Witten]. 
The precise relation between the dynamics of the
two is the subject of a forthcoming paper [\HLWnew].
In the next
section we will obtain multi-threebrane  solutions and describe their
elementary properties. In the final section we will use the threebrane
solitons to obtain $(p-1)$-branes and $(p-2)$-branes on the D-$p$-branes 
of type II string theory.


\chapter{The Threebrane Soliton}

In this paper we denote the six-dimensional coordinates of the fivebrane by 
hatted variables $\m,\n = 0,1,2,...,5$. In addition all indices are raised 
and lowered with respect to the flat Euclidean metric unless explicitly 
indicated otherwise.
Let us look for a threebrane in the plane $x^0,x^1,x^2,x^3$. We let  
unhatted variables refer to the transverse
coordinates of the threebrane i.e. $x^4,x^5$. We will assume all fields to
depend only on the transverse coordinates.
To find a threebrane we dualise one scalar $X^{1'}$ to a five-form $G_5$
$$
G_{\m\n\p\q\r} =  \epsilon_{\m\n\p\q\r\s}\partial^{\s}X^{1'} \ .
\eqn\Gdef
$$
Our ansatz is then
$$\eqalign{
G_{0123m} & =\epsilon_{mn}\partial^n X^{1'} \equiv v_m \ ,\cr
X^{2'} &=\phi\ ,\cr
h_{mnp}&=0\ ,\cr}
\eqn\ansatz
$$ 
where $\epsilon_{mn}$ is the volume element on the transverse space and 
the other scalars $X^{3'},X^{4'},X^{5'}$ are constant. With this
ansatz the equations of motion of the fivebrane simplify greatly. We introduce
the metric
$$
g_{\m\n} =\left(\matrix{
-1&0&\cr
0&{\bf 1}_{3\times 3}&\cr
&&\delta_{mn} + \partial_{m}X^{1'}\partial_{n}X^{1'}+ 
\partial_{m}X^{2'}\partial_{n}X^{2'}\cr
}\right) \ ,
\eqn\gdef
$$
where ${\bf 1}_{3\times 3}$ this the unit matrix in three dimensions. Since 
the three form is zero the only equations of motion we need to solve for 
are [\HSW]
$$
g^{mn}\nabla_m\nabla_nX^{1'} = g^{mn}\nabla_m\nabla_nX^{2'}=0 \ ,
\eqn\eqom
$$
where $\nabla$ is the Levi-Civita connection of $g_{\m\n}$.

As with the string soliton found in [\HLW] it is helpful now to determine
the condition that half of the supersymmetry is preserved by the soliton. 
It is instructive to first consider the linearised supersymmetry
$$\eqalign{
\delta_0\Theta_{\beta}^{\ j} &= \epsilon^{\alpha i}\left({1\over2}
(\gamma^m)_{\alpha\beta}(\gamma_{b'})_{i}^{\ j}\partial_m X^{b'}
-{1\over6}(\gamma^{mnp})_{\alpha\beta}\delta_i^{\ j}h_{mnp}\right) \ ,\cr
&= {1\over2}\epsilon^{\alpha i}
(\gamma^m)_{\alpha\beta}(\gamma_{b'})_{i}^{\ j}\partial_m X^{b'}\ ,\cr}
\eqn\linsusy
$$
for $h_{mnp}=0$.
Inserting our ansatz in to $\delta_0\Theta_{\beta}^{\ j} =0$ one finds
$$\eqalign{
0={1\over2}\epsilon^{\alpha i}(\gamma^4)_{\alpha\gamma}(\gamma_{1'})_{i}^{\ k}
& \left\{\left[
\delta_{k}^{\ j}\delta^{\gamma}_{\ \beta} v_5 
+ (\gamma_{1'2'})_{k}^{\ j}(\gamma^{45})^{\gamma}_{\ \beta}\partial_5\phi
\right] \right. \cr
&\left.
+ (\gamma_{1'2'})_{k}^{\ l} 
\left[
\delta_{l}^{\ j}\delta^{\gamma}_{\ \beta}\partial_4\phi 
+ (\gamma_{1'2'})_{l}^{\ j}(\gamma^{45})^{\gamma}_{\ \beta}v_4
\right]\right\}\ .\cr}
\eqn\linsusytwo
$$
Thus if we set
$$
v_m = \pm\partial_m\phi \ ,
\eqn\bogocond
$$
we find that the solution will be invariant under supersymmetries 
which satisfy
$$
\epsilon_0^{\alpha i}
(\gamma_{1'2'})_i^{\ j}(\gamma^{45})_{\alpha}^{\ \beta}
= \mp\epsilon_0^{\beta j} \ .
\eqn\ressusy
$$
Note that  the bogomol'nyi condition \bogocond\ is equivalent to the 
statement that 
$X^{1'}+iX^{2'}$ is an (anti-)holomorphic function of 
$x^4+ix^5$ for the
minus (plus) sign in \bogocond .
To verify  that the supersymmetries are preserved to all orders  we recall the 
expression obtained in [\HLW] for the full non-linear 
supersymmetry
$$
\delta\Theta^{\ j}_{\beta} = 
-{1\over2}{1\over\sqrt{-\det g}}
\epsilon^{\alpha i}(E^{-1})_{\alpha i}^{\ \ \gamma k}
\partial_m X^{b'}(\gamma^m)_{\gamma\delta}(\gamma_{b'})_{k}^{\ l}
(u)_{\ l\beta}^{\delta \ \ j}\ .
\eqn\fullsusy
$$
Here we have set $h_{mnp}=0$ and 
we refer the reader to [\HLW] for a detailed description of the 
matrices $E$ and $u$. 
Clearly \fullsusy\ vanishes if $\delta_0\Theta_{\beta}^{\ j}=0$, provided that 
the preserved supersymmetries are $\epsilon^{\beta j} = 
\epsilon_0^{\alpha i}E_{\alpha i}^{\ \ \beta j}$, where 
$\epsilon_0^{\alpha i}$ satisfies \ressusy . 

If we substitute the Bogomol'nyi condition \bogocond\ into the metric
\gdef\ it is now an easy matter to see that the equation of motion becomes
$$
{1\over 1 + (\partial\phi)^2}\partial^2\phi = 0 \ ,
\eqn\laplace
$$
where $\partial^2\phi =\delta^{mn}\partial_m\partial_n\phi$ and
$(\partial\phi)^2=\delta^{mn}\partial_m\phi\partial_n\phi$. 
Furthermore it follows from 
the Bogomoln'yi condition \bogocond, that  \laplace\ is solved
by any $X^{1'},X^{2'}$.
One can also check from \laplace\ that $G_5$ is a closed form. Thus
the solution corresponding to $N$ threebranes located at $y_I$
($I=0,\ldots, N-1$)
has the general form
$$
\eqalign{
G_{0123m} & = \pm \partial_m\phi\ ,\cr
\phi & = \phi_0 + \sum_{I=0}^{N-1}Q_I\ln |x - y_I| \ , \cr}
\eqn\solution
$$
where 
$\phi_0$ and $Q_I$ are constants. 
Clearly this solution has bad asymptotic behaviour unless $\sum Q_I=0$,
however one can still define the charge of a single threebrane 
$$
Q = {1\over 2\pi}\int_{S_{\infty}^1} \star G_5 = \pm Q_0\ ,
\eqn\Gcharge
$$
where $S^1_{\infty}$ is the transverse circle at infinity and $\star$ is the 
flat six-dimensional Hodge star.
The presence of the conformal factor in \laplace\ indicates that 
the equations of motion are satisfied even at the points where
the solution is ill behaved and hence no sources are needed. 

Let us consider the zero modes of a single threebrane 
soliton. Clearly there are two bosonic
zero modes $y^m_0$ describing the location of the threebrane in the
transverse space. Now consider the three-form 
$h_3 = db_2 = 0$. The closed two form $b_2$ has a gauge symmetry
$b_2\rightarrow b_2 + dA_1$. However, because of  Poincar\'e invariance
of the threebrane along $x^0,x^1,x^2,x^3$, there are vector fields $A_1$
whose indices tangent to the threebrane do not vanish at infinity, 
corresponding to so-called
large gauge transformations.
Following the standard treatment of soliton zero-modes 
(see for example [\Harvey]),
we must interpret these components of $A_1$ as  zero modes of the 
threebrane. Thus there is also a four dimensional
vector zero mode living on the threebrane worldvolume.\foot{The same reasoning
also implies that there is a two dimensional vector zero mode on the
self-dual string soliton worldvolume. This was ignored in [\HLW] because 
it carries no degrees of freedom.} The fermionic
zero modes come from the broken supersymmetries and hence there are eight of 
them and in four dimensions they are necessarily non-chiral.
Thus we find an $N=2$, $D=4$ vector multiplet of zero modes on the
threebrane.

Since the scalars transform under $Spin(5)$, by choosing an arbitrary 
pair of  scalars
$(X^{a'}, X^{b'})$ with $a'\ne b'$ for our solution, we obtain a multiplet of 
threebranes transforming as a ${5 \times4\over 2}=10$ dimensional
representation of $Spin(5)$, in agreement with the algebra \algebra .
In the  
M theory interpretation, where the threebrane represents the intersection
of two fivebranes, these scalars 
point along the two directions of the external fivebrane which are 
transverse to the worldvolume of the fivebrane we are considering.

We can also see that there is no BPS instanton configuration
or $-1$-brane. For in this case we would again have only two non-trivial 
scalars, $X^{1'}$ and $X^{2'}$, with $\partial_m X^{1'}$ interpreted as the
field strength of the $-1$-brane. However, what made the threebrane 
supersymmetric was the fact that in the two-dimensional transverse space 
the two scalars (or more precisely their field strengths) are dual to each 
other. For a $-1$-brane, where the transverse space is the six-dimensional 
Euclidean  worldvolume of the fivebrane, there is no such condition. In 
fact one could
always perform a $Spin(5)$ rotation to a configuration
with just one scalar. Clearly such a configuration could not preserve
any supersymmetry.


\chapter{Discussion}

In this paper we have discussed the one half supersymmetric BPS states of
the M theory fivebrane and in particular we obtained a new threebrane soliton.
We also determined the zero modes of resulting $p$-branes. We note that
the zero modes can be 
obtained by the dimensional reduction of a $D=6$, $N=(1,0)$ super-Maxwell 
multiplet to $p+1$ dimensions. We would like to conclude with a brief 
discussion of the dimensional reduction of the threebrane to D-$p$-brane
worldvolumes.

It was shown in [\HSW] that the  fivebrane equations of motion can
be dimensionally reduced to those of the D-fourbrane. Thus the
threebrane soliton can be double dimensionally reduced to  a twobrane 
on the D-fourbrane. By T-duality
these can be extended to $(p-2)$ solitons on a D-$p$-brane
(we will restrict our attention to $p\ge2 $).
These $(p-2)$-branes have exactly the same form as \solution , only with
$G_5$ now replaced by a $p$-form $G_{0\ldots (p-2) m}=\pm\partial_m\phi$. 
These solutions represent two intersecting D-$p$-branes
over a $(p-2)$-brane. 

The above discussion only works for $p<8$ since the D-eightbrane only has 
one scalar field on its worldvolume. Therefore the naive extrapolation
of the soliton \solution\ which involves two scalars cannot work. Instead
the vector field on the eightbrane must become non-zero. More precisely
consider two D-fourbranes  intersecting over a 
twobrane (say one in the $x^0,x^1,x^2,x^3,x^4$ plane
and the other in the $x^0,x^1,x^2,x^5,x^6$ plane). Now T-dualise along
$x^5,x^6,x^7,x^8$ to obtain an eightbrane in the $x^0,\ldots,x^8$ 
plane, containing
a fourbrane in the $x^0,x^1,x^2,x^7,x^8$ plane. 
We expect that this configuration is actually described by an instanton
in the four transverse dimensions to the fourbrane; $x^3,x^4,x^5,x^6$. 
To see this explicitly let us construct the vector field $A_m$ in 
this transverse space. Now, in order to have applied T-duality we must 
set $A_{5}=A_6=0$ (or at least they must be pure gauge). 
Consequently, if we reinterpret 
$A_3 = X^{1'}$ and $A_4 = X^{2'}$, we find that the Bogomol'nyi 
condition \bogocond\ (i.e. holomorphicity) is equivalent to 
self-duality of the gauge field $A_m$.

One could also directly dimensionally reduce the threebrane soliton \solution\
to a threebrane on the D-fourbrane worldvolume. 
This leads to a  domain wall on the fourbrane and is  
associated with a cosomological constant, 
which can be obtained by a Sherk-Schwarz type 
of dimensional reduction invoking one of the scalars. 
The form of this solution should also be clear
to the reader, the only  modification being that there is only one transverse
dimension so that the logarithms in \solution\ are 
replaced by linear functions.
This threebrane soliton represents a D-fourbrane intersecting with an 
NS-fivebrane. Clearly this solution can also be T-dualised to  other
D-$p$-branes intersecting with an NS-fivebrane over a $(p-1)$-brane.

Similarly to the above case this only works for $p<7$. To obtain a sevenbrane
one must T-dualise in the transverse space of the NS-fivebrane. This leads
to a Kaluza-Klein like fivebrane (i.e. a geometry which is 
${\bf R}^6\times M_{Taub-NUT}$),  bound
to the sevenbrane, which is wrapped around one compact and one non-compact
dimension of $M_{Taub-NUT}$. It is
impossible to T-dualise again to obtain an eightbrane configuration, since 
there is no suitable Taub-Nut geometry with the (local) form 
${\bf R}^2\times {\bf T}^2$. This is reflected by the fact that there is no
way to obtain a cosmological constant in the nine-dimensional Maxwell action 
by performing a 
Sherk-Schwarz reduction from ten dimensions.
 
We would like to thank G. Papadopolous and H.A. Chamblin for discussions.

\refout

\end